\documentclass[12pt]{article}

\usepackage{amsmath}
\usepackage{amssymb}
\usepackage{graphicx}
\usepackage{float}
\usepackage[dvips]{color}
\usepackage{colordvi}
\usepackage{url}
\usepackage{wrapfig}

\usepackage{comment}
\input{colordvi.tex}

\allowdisplaybreaks[1]
\newcommand{\nt}{\nonumber\\}

\newcommand{\bbZ}{{\mathbb Z}}

\newcommand{\cF}{{\cal F}}

\newcommand{\cL}{{\cal L}}

\newcommand{\cN}{{\cal N}}

\newcommand{\bbC}{{\mathbb C}}

\newcommand{\Tr}{{\rm Tr}}

\newcommand{\dg}{\dagger}

\newcommand{\ba}{\begin{eqnarray}}
\newcommand{\ea}{\end{eqnarray}}
\newcommand{\bag}{\begin{align}}
\newcommand{\eag}{\end{align}}
\newcommand{\sss}[1]{\subsubsection*{#1}}

\newcommand{\qqquad}{\qquad\qquad}

\newcommand{\eps}{\epsilon}

\begin{document}

\thispagestyle{empty}
\addtocounter{page}{-1}
{}
\vskip-5cm
\begin{flushright}
KEK-TH-1628
\end{flushright}
\vspace*{0.7cm} \centerline{\Large \bf 
Thermodynamics of black M-branes from SCFTs}
\vspace*{1 cm} 
\centerline{\bf 
 Takeshi~Morita and Shotaro Shiba}
\vspace*{0.7cm}
\centerline{\rm \it KEK Theory Center}
\vspace*{0.1cm}
\centerline{\it High Energy Accelerator Research Organization (KEK)}
\vspace*{0.1cm}
\centerline{\it 1-1 Oho, Tsukuba, Ibaraki 305-0801, \rm JAPAN}
\vspace*{0.5cm}
\centerline{\tt E-mail:{\!\!} tmorita(at)post.kek.jp, sshiba(at)post.kek.jp}
\vspace*{1cm}
\centerline{\bf Abstract}
\vspace*{0.5cm} 
\enlargethispage{1000pt}

We discuss thermodynamics of $N$ M2-branes at 
strong coupling 
from the ABJM theory by employing the Smilga-Wiseman method, which explains the black D$p$-brane thermodynamics from the maximally supersymmetric $U(N)$ Yang-Mills theories through a field theory analysis.
As a result we obtain the free energy of the ABJM theory $\sim N^{3/2} \sqrt{k} T^3$, which is consistent with the prediction from the eleven-dimensional supergravity.
We also estimate the free energy of $N$ M5-branes 
by assuming some natural properties of the 6d superconformal field theory.
Remarkably we obtain the free energy $\sim N^3 T^6$, which is consistent again with the supergravity prediction.
This result might illuminate the low energy field theory description of the multiple M5-branes.

\newpage

\section{Introduction}

Understanding the $N$-dependence of the thermal free energies of $N$ M2 and M5-branes is an outstanding problem in string theory.
They are predicted through the eleven-dimensional supergravity to be proportional to $N^{3/2} T^3$ and $N^{3} T^6$, respectively~\cite{Klebanov:1996}.
They are contrasting to the free energies of $N$ D-branes, which are always proportional to $N^2$ in the string theory regime. Then we have believed that those natures would be deeply connected to the non-perturbative aspects of the mysterious M-brane theories.

In the case of $N$ D$p$-branes, while the $N$-dependence is simple, the supergravity analysis predicts a non-trivial dependence on the 't Hooft coupling as \cite{Itzhaki:1998dd}
\begin{align}
F \sim N^2 T^{\frac{2(7-p)}{5-p}} \lambda^{-\frac{3-p}{5-p}} V_p\,,
\end{align}
where $T$ is the temperature, $V_p$ is the spatial volume of the D$p$-brane and
 $\lambda=g_{YM}^2N$ is the 't Hooft coupling of the $p+1$ dimensional maximally supersymmetric Yang-Mills theory.
This gravity result is reliable when $N^{-\frac{2}{7-p}} \ll  (T \lambda^{-\frac{1}{3-p}} )^{\frac{3-p}{5-p}} \ll 1$ is satisfied.
Recently, Wiseman~\cite{Wiseman:2013} has successfully
explained
this result for $p<3$ cases from a gauge theory calculation by generalizing Smilga's method for D0-brane~\cite{Smilga:2008}.\footnote{ The Smilga's method
\cite{Smilga:2008}
agrees with numerical studies in the D0-brane case \cite{Anagnostopoulos:2007fw} and the bosonic BFSS case \cite{Aharony:2004ig, Kawahara:2007fn}. In the bosonic BFSS case, we can explicitly confirm it through an analytical computation \cite{Mandal:2009vz}. 
}
In addition, the size of the horizon is also correctly estimated.
In this derivation, the classical moduli fields play  important roles at low temperature.
We call this method as Smilga-Wiseman method in this paper.

This success encourages us to investigate the thermodynamics of M-branes through a similar approach.
In the case of $N$ M2-branes, the ABJM theory is known as the low energy field theory~\cite{Aharony:2008}. 
Then we apply the Smilga-Wiseman method to this theory, and show the free energy $F \sim N^{3/2} \sqrt{k} T^3$, which is consistent with the supergravity prediction.

In the case of $N$ M5-branes, a 6d $\cN=(2,0)$ superconformal field theory would describe the low energy dynamics \cite{Maldacena:1997re}, but the details of this theory have not been uncovered yet.
Thus we cannot derive the free energy directly from the effective theory.
In this paper,
 instead of considering the whole effective theory,
 we assume some natural properties of the classical moduli fields of this theory and estimate the free energy through the Smilga-Wiseman method. 
 Then we indeed obtain $F \sim N^{3} T^6$, which is consistent with the supergravity prediction.
Therefore this result may give us some clues to understand the low energy dynamics of multiple M5-branes.

This paper is organized as follows.
In section~\ref{sec:Dp}, we review the Smilga-Wiseman method and the discussion on the D$p$-brane case. 
By using this method, in section~\ref{sec:M2} we study the thermodynamics of M2-branes in the ABJM theory.
We also study the thermodynamics of M5-branes in section~\ref{sec:M5}, making some natural assumptions on the effective M5-brane theory.
In section~\ref{sec:argue}, we compare the M and D-brane results and argue why the exotic $N$-dependence appears in the M-branes.
Discussions are in section~\ref{sec:Disc}.
The corresponding results from the dual supergravity are summarized in appendix~\ref{sec:gr}.
In appendix~\ref{sec:app}, we mention the details of calculation on the one-loop quantum corrections. 

\section{Smilga-Wiseman method for D$p$-branes}
\label{sec:Dp}

We demonstrate how Smilga and Wiseman estimated the free energy of the black D$p$-branes ($p<3$) from the $p+1$ dimensional maximally supersymmetric U($N$) Yang-Mills theory (SYM) at low temperature~\cite{Wiseman:2013,Smilga:2008}. 

It is convenient to use the Euclidean time formulation at finite temperature $T$, where Euclidean time $\tau=ix^0$ is periodic with a period of $\beta=1/T$.  Then the Euclidean SYM action is
\bag
S_{\text{D}p}&=
\frac{N}{\lambda}\int d\tau d^px \,\Tr\left[\frac14 \cF_{\mu\nu}^2
+\frac12(D_\mu\Phi^I)^2+\frac{i}{2}\bar\Psi\Gamma^\mu D_\mu\Psi
\right.\nt
&\left.\qqquad\qqquad\qquad~
-\frac14[\Phi^I,\Phi^J]^2-\frac{i}{2}\bar\Psi\Gamma^I[\Phi^I,\Psi]\right],
\label{action-Dp}
\end{align}
where $\mu=0,\ldots,p$ and $I=1,\ldots,9-p$.
A set of classical vacua of this theory is gauge-equivalent to the configurations
\ba\label{Dp-vacua}
A_{\mu,ab}=a_{\mu,a}\delta_{ab}\,,\quad
\Phi^I_{ab}=\phi^I_a\delta_{ab}\,,\quad
\Psi_{ab}=0\,,
\ea
where $a,b=1,\ldots, N$ are the indices of the adjoint hermitian matrices. $a_{\mu,a}$ and $\phi_a^I$ are real constants, and $\phi_a^I$ represent the positions of the D$p$-branes in the transverse directions.
Note that this configuration breaks the original $U(N)$ gauge symmetry to $U(1)^N$.

At zero temperature, $\phi^I_a$ are the exact moduli and thus each D$p$-brane can stand at any place.
At low temperature, in the following arguments, we assume that the dominant configurations of the D$p$-branes satisfy 
\begin{align}
 \beta |\phi_a-\phi_b| \gg 1\,,
 \label{cond-long}
 \end{align}
where $|\phi_a-\phi_b|:=\sqrt{\sum_I(\phi_a^I-\phi_b^I)^2}$.
In this case, 
the off-diagonal components of the scalars, fermions and gauge fields possess large mass $\sim |\phi_a-\phi_b|$, so they can be integrated out. Then the effective theory of the moduli  $\phi^I_a$ would be relevant to determine the low temperature dynamics.\footnote{
We assume that the moduli from the gauge field $a_{\mu,a}$ do not play relevant roles in the D$p$-brane cases. In the M2-brane case, this condition would be ensured since only Chern-Simons gauge fields exist in the ABJM theory.
In the M5-brane case, we assume again this condition although the kinematics of the self-dual 2-form gauge field in 6d $\cN=(2,0)$ theory has not been clearly understood.
}
We will see later that the condition $\beta |\phi_a-\phi_b| \gg 1$ is self-consistently satisfied.

Now we consider the effective theory of the moduli $\phi^I_a$ at low temperature.
From the classical action (\ref{action-Dp}), we obtain the classical term
\begin{align}
S_{\text{D}p}^{\text{classical}} = \frac{N}{\lambda} \int d \tau d^px \sum_{a} \left( \frac{1}{2} \partial^\mu \phi^I_a \partial_\mu \phi^I_a 
\right).
\label{action_cl}
\end{align}
In addition,
 quantum corrections are induced through the integrals of the massive fields.
Here we just consider the leading one-loop correction.\footnote{We will comment on the higher loop corrections in section~\ref{sec:Disc}. }
We distinguish between the temperature independent corrections, which do not depend on temperature explicitly, and the temperature dependent terms.
At the one-loop level, the temperature independent potential is given by
\begin{align}
S_{\text{D}p,T=0}^{\text{one-loop}} &= -\int d \tau d^px \sum_{a<b} \frac{\Gamma\left(\frac{7-p}{2}\right)}{(4\pi)^\frac{1+p}{2}}\left(  
 2\frac{\{\partial_\mu (\phi_a^I -\phi_b^I) \partial_\nu (\phi_a^I -\phi_b^I)\}^2}
 {|\phi_a -\phi_b|^{7-p}} \right. \nt
& \left.\qqquad\qqquad\quad~
-\frac{ \{ \partial_\mu  (\phi_a^I -\phi_b^I) \partial^\mu (\phi_a^I -\phi_b^I)\}^2 }{|\phi_a -\phi_b|^{7-p}} 
\right) + \ldots\,.
\label{action_one}
\end{align}
Here `$\ldots$' denotes the subdominant terms which are suppressed when the distances between 
the branes are large and $ | (\partial \phi)^2  /\phi^4| \ll 1$ is satisfied.
This potential is attractive at a long distance, and so force to confine the D$p$-branes.
An important point is that this quantum correction starts from the quartic derivative terms $(\partial \phi)^4$ since the supersymmetry protects the quadratic terms $(\partial \phi)^2$ from the corrections. 

The effective theory has the temperature dependent terms also.
However, as argued in appendix~\ref{sec:app}, these terms are proportional to
$\exp(-\beta|\phi_a -\phi_b|)$, and so strongly suppressed under the condition (\ref{cond-long}). 

Therefore the relevant terms at low temperature at the one-loop level are the classical term (\ref{action_cl}) and the leading term of the temperature independent potential (\ref{action_one}).
Then Smilga and Wiseman 
argued that the loop expansion is strongly coupled in the gravity regime and the classical moduli action (\ref{action_cl}) and  one-loop potential (\ref{action_one}) are balanced. 
Presumably this may be related to the virial theorem.
Thus we estimate the order of the moduli $\phi_a^I$ by equating these two terms.

In this configuration, we estimate that the moduli $\phi_a^I$ and their differences are the same order which we denote as $\phi$\,:
\begin{align}
\phi^I_a & \sim \phi^I_a-\phi^I_b \sim \phi\,.
\label{phi-same}
\end{align}
To determine the scale $\phi$, we 
assume that the derivative terms satisfy
\begin{align}
\partial \phi^I_a & \sim \partial (\phi^I_a-\phi^I_b) \sim \frac{1}{\beta} \phi\,.
\label{kin-beta}
\end{align}
This is a key assumption of the Smilga-Wiseman method.
This implies that the temperature dominates the moduli dynamics rather than the 't Hooft coupling $\lambda$, which is another dimensionful parameter of the theory. 
This may be natural from the view of the classical moduli.

By using the assumptions (\ref{phi-same}) and (\ref{kin-beta}), the effective action (\ref{action_cl}) and (\ref{action_one}) can be estimated as
\begin{align}
S_{\text{D}p}^{\text{classical}}  \sim  \int d^px \frac{N^2}{\beta \lambda} \phi^2\,, \quad
S_{\text{D}p,T=0}^{\text{one-loop}} \sim \int  d^px  \frac{N^2}{\beta^3 \phi^{3-p}}\,,
\end{align}
where we have used $\int d\tau \sim \beta$\,, $\sum_a \sim N$ and $\sum_{a<b} \sim N^2$. 
Therefore if these two terms are balanced
\ba
S_{\text{D}p}^{\text{classical}}  \sim S_{\text{D}p,T=0}^{\text{one-loop}}\,,
\ea
we obtain the scale of the scalar fields as 
\begin{align}
\phi \sim T^{\frac{2}{5-p}} \lambda^{\frac{1}{5-p}}\,.
\label{phi-scale}
\end{align}
This result is consistent with the scale estimated from the size of the horizon of the black D$p$-brane (\ref{size-Dp}).
Note that the condition (\ref{cond-long}) 
is satisfied at low temperature
$  (T \lambda^{-\frac{1}{3-p}} )^{\frac{3-p}{5-p}} \ll 1$ where the gravity description is reliable, and the assumption that the thermal loop corrections are suppressed is justified self-consistently.

By substituting 
this value into the action $S_{\text{D}p}^{\text{classical}}  \sim S_{\text{D}p,T=0}^{\text{one-loop}} \sim S_{\text{D}p}$, we obtain the free energy as 
\begin{align}
F_{\text{D}p} \sim S_{\text{D}p}/\beta \sim  N^2 T^{\frac{2(7-p)}{5-p}} \lambda^{-\frac{3-p}{5-p}} V_p\,,
\label{F-Dp}
\end{align} 
where $V_p$ is the spatial volume of the D${p}$-brane. 
This result reproduces the parametric dependence of 
the free energy of the black D$p$-brane from the dual gravity (\ref{F-Dp-gr}).

Note that although Wiseman focused on $p<3$ cases, 
this derivation is applicable to $p=3,4$ and 6.\footnote{In $p=5$ case, we obtain no conditions on the scale $\phi$.}
As far as the temperature satisfies $  (T \lambda^{-\frac{1}{3-p}} )^{\frac{3-p}{5-p}} \ll 1$, the obtained scale $\phi$ (\ref{phi-scale}) satisfies the condition (\ref{cond-long}) and the assumption that the thermal corrections are suppressed is justified.
Interestingly, this condition $  (T \lambda^{-\frac{1}{3-p}} )^{\frac{3-p}{5-p}} \ll 1$ is equivalent to the small curvature condition in the gravity \cite{Itzhaki:1998dd}.

\section{ABJM theory}
\label{sec:M2}

The ABJM theory is the three-dimensional $\cN=6$ $U(N)\times U(N)$ Chern-Simons gauge theory with level $k$ and $-k$, which is dual to M-theory on $AdS_4\times S^7/\bbZ_k$ \cite{Aharony:2008}. It is known as the effective field theory on $N$ M2-branes.

Now we apply the Smilga-Wiseman method to this theory in order to discuss the thermodynamics of M2-brane system.
The Euclideanized ABJM action is 
\bag
S_\text{ABJM}&=\frac{k}{2\pi}\int d\tau d^2x\,\left(
\Tr\left[
(D_\mu\Phi_A^\dagger)(D^\mu \Phi^A)
+i\Psi^{\dagger A}\gamma^\mu D_\mu \Psi_A
\right]\right.
\nt&\left.\qqquad\qquad~~~~
+\cL_\text{CS}^{(1)}-\cL_\text{CS}^{(2)}-V_B-V_F\right),
\label{action-ABJM}
\end{align}
where 
\bag
\cL_\text{CS}^{(i)}&=\frac12\eps^{\mu\nu\rho}\,\Tr\left[A^{(i)}_\mu\partial_\nu A^{(i)}_\rho+\frac{2}{3}A_\mu^{(i)}A_\nu^{(i)}A_\rho^{(i)}\right],
\nt
V_B&=
\frac13\Tr\left[\Phi_A^\dg\Phi^A\Phi_B^\dg\Phi^B\Phi_C^\dg\Phi^C
+\Phi^A\Phi_A^\dg\Phi^B\Phi_B^\dg\Phi^C\Phi_C^\dg
+4\Phi^A\Phi_B^\dg\Phi^C\Phi_A^\dg\Phi^B\Phi_C^\dg
\right.
\nt&\left.\qquad\quad~
-6\Phi^A\Phi_B^\dg\Phi^B\Phi_A^\dg\Phi^C\Phi_C^\dg\right],
\nt
V_F&=
i\Tr\left[\Phi_A^\dg\Phi^A\Psi^{\dg B}\Psi_B
-\Phi^A\Phi_A^\dg\Psi_B\Psi^{\dg B}
-2\Phi_A^\dg\Phi^B\Psi^{\dg A}\Psi_B
+2\Phi^A\Phi_B^\dg\Psi_A\Psi^{\dg B}
\right.
\nt&\left.\qquad~~~
-\eps^{ABCD}\Phi_A^\dg\Psi_B\Phi_C^\dg\Psi_D
+\eps_{ABCD}\Phi^A\Psi^{\dg B}\Phi^C\Psi^{\dg D}\right],
\end{align}
where $A,B,C,D=1,\ldots,4$, which denote the transverse directions in $\bbC^4$ (orbifolded by $\bbZ_k$).
The matter fields are complex scalar fields $\Phi^A$ and spinor fields $\Psi_A$ which transform as ${\bf 4}$ and ${\bf \bar 4}$ under $SU(4)$ R-symmetry, respectively.
Both fields have bifundamental representations under the $U(N)\times U(N)$ gauge group.
The covariant derivative is defined as
$D_\mu\Phi^A = \partial_\mu \Phi^A + iA^{(1)}_\mu \Phi^A - i\Phi^A A_\mu^{(2)}$, {\it etc.}

Let us focus on the dynamics of the moduli fields.
The configuration of the classical vacua which we consider here is 
\ba
\Phi^A_{ab} = \frac{1}{\sqrt{2}}(\phi^A_a + i\phi^{A+4}_a)\delta_{ab}\,,\quad
\Psi_{ab} = 0\,,
\ea
where $\phi^I_a$ are real scalar fields ($I=1,\ldots,8$).
Then the moduli of the ABJM theory are given by the diagonal components $\phi^I_a$ of the bifundamental scalars $\Phi^A_{ab}$. 
The classical action is given by
\begin{align}
S_{\text{ABJM}}^{\text{classical}} = \frac{k}{2\pi} \int d \tau d^2x \sum_{a} \left( \frac{1}{2} \partial^\mu \phi^I_a \partial_\mu \phi^I_a \right).
\label{ABJM_cl}
\end{align}
The one-loop effective potential for the moduli at zero temperature has been calculated in \cite{Baek:2008},
\begin{align}
S_{\text{ABJM},T=0}^{\text{one-loop}} &= -\int d \tau d^2x \sum_{a<b} \frac{5}{16\pi}\left(  
 2\frac{\{\partial_\mu (\phi_a^I -\phi_b^I) \partial_\nu (\phi_a^I -\phi_b^I)\}^2}
 {|\phi_a -\phi_b|^{6}} \right. \nt
& \left.\qqquad\qqquad~~
-\frac{ \{ \partial_\mu  (\phi_a^I -\phi_b^I) \partial^\mu (\phi_a^I -\phi_b^I)\}^2 }{|\phi_a -\phi_b|^{6}} 
\right) + \ldots\,.
\label{ABJM_one}
\end{align}
The quantum correction starts from the quartic derivative terms 
$(\partial \phi )^4$, just as in the D$p$-brane case (\ref{action_one}).
It is consistent with the gravitational attractive potential at a long distance.

The temperature dependent terms can be estimated as in appendix~\ref{sec:app},
and should be proportional to 
$\exp(-\beta|\phi_a-\phi_b|^2)$,
so they are suppressed when the condition
$\beta |\phi_a -\phi_b|^2 \gg 1$ is satisfied.

Then by imposing the assumptions (\ref{phi-same}) and (\ref{kin-beta}) as in the D$p$-brane case, we estimate the actions as
\begin{align}
S_{\text{ABJM}}^{\text{classical}}  \sim  \int d^2 x \frac{kN}{\beta } \phi^2\,, \quad
S_{\text{ABJM},T=0}^{\text{one-loop}} \sim \int  d^2 x  \frac{N^2}{\beta^3 \phi^{2}}\,.
\end{align}
By equating 
$S_{\text{ABJM}}^{\text{classical}}  \sim S_{\text{ABJM},T=0}^{\text{one-loop}}$, 
we  obtain 
\begin{align}
\phi \sim \frac{N^{\frac14}}{k^{\frac14} \beta^{\frac12}}\, ,
\end{align}
which 
is consistent with the condition $\beta \phi^2 \gg 1$ at large $N$ (with $k$ fixed).
This size agrees with the dual gravity prediction (\ref{size-M2}).

Now we insert this result to the action $S_{\text{ABJM}}^{\text{classical}}  \sim S_{\text{ABJM},T=0}^{\text{one-loop}} \sim S_{\text{ABJM}}$, and obtain
\begin{align}
F_{\text{ABJM}} \sim S_{\text{ABJM}}/\beta \sim N^{\frac32} \sqrt{k} T^3 V_2\,.
\label{free-ene-ABJM}
\end{align}
This estimate reproduces the dual gravity prediction (\ref{F-ABJM}).

Note that if we replace $k$ with the 't Hooft coupling $\lambda=N/k$, the free energy (\ref{free-ene-ABJM}) becomes 
\begin{align}
F_{\text{ABJM}} \sim N^{\frac32} \sqrt{k} T^3 V_2 \sim N^2\lambda^{-\frac12}T^3V_2 \sim F_{\text{IIA}}\,.
\label{M2-D2}
\end{align}
The right hand side is the free energy of the system in the IIA supergravity regime~\cite{Aharony:2008}.
Thus the M and string theory would be connected smoothly not only in the perspective of the gravity as argued in \cite{Aharony:2008} but also in the gauge theory.\footnote{
The free energy of D2-branes $\sim N^2\lambda^{-\frac13}T^{\frac{10}{3}}V_2$ in eq.\,(\ref{F-Dp}) is not smoothly connected to that of M2-branes (\ref{free-ene-ABJM}) with $k=1$, since a Gregory-Laflamme
 transition would occur \cite{Itzhaki:1998dd, Gregory:1994, Martinec:1998ja}.
It must be interesting if we can see this transition in the ABJM theory.}

In gauge theories, the exotic $N^{3/2}$ dependence of the free energy of M2-branes had been derived only through the localization technique~\cite{Drukker:2010,Fuji:2011,Hanada:2012}, although it is not a thermal one.
It is surprising that the Smilga-Wiseman method explains the $N$ dependence via much simpler estimation.

\section{6d superconformal field theory}
\label{sec:M5}

We estimate the thermodynamics of M5-brane system from the 6d superconformal field theory.
The low energy dynamics of $N$ M5-branes is described by the 6d $\cN=(2,0)$ SCFT,
but the precise formulation of this theory has not been found.
Therefore, in our analysis, we make the following assumptions on the effective theory of M5-branes:

\begin{enumerate}
\item 
The theory has the scalar moduli fields $\phi^I_a$ ($I=1,\ldots, 5$ and $a=1,\ldots,N$) which represent the collective motion of $N$ M5-branes in the transverse directions.
\item
The  effective potential for the moduli $\phi_a^I$ starts from $ (\partial \phi )^4 $ for a small $\partial \phi$ and a long distance $|\phi_a-\phi_b|$. \end{enumerate}

The first assumption must be a natural requirement, since M5-brane is a BPS object.
The second assumption comes from the observation in the D$p$ and M2-brane cases at long distances.
When M5-branes are separated from each other, massive fields may be integrated out and we may obtain an effective action for the moduli fields.
Then similar discussion to the D$p$ and M2-brane cases can be done:
The quadratic derivative terms of the moduli $(\partial\phi)^2$ should not receive the quantum correction due to the supersymmetry, and so the correction should start from the quartic derivative terms $(\partial\phi)^4$.
In addition to this potential, there would be thermal corrections, which depend on temperature explicitly.
However, as argued in appendix~\ref{sec:app},
they would be suppressed by $\exp(-\beta \sqrt{ |\phi_a-\phi_b| })$ and we can ignore them when the condition $\beta \sqrt{ |\phi_a-\phi_b|}  \gg 1 $ is satisfied.

By using these assumptions, we estimate the effective action for the scalar moduli $\phi^I_a$.
The kinetic term can be read off from the single M5-brane action \cite{PST,Bandos:1997, Cederwall:1998}\footnote{The M5-brane has the self-dual two form fields and fermion fields too.
 But we ignore them and focus on the scalar moduli dynamics.
}, and we obtain
\begin{align}
S_{\text{M5}}^{\text{kin}} \sim \int d \tau d^5x \sum_{a} \left( \frac{1}{2} \partial^\mu \phi^I_a \partial_\mu \phi^I_a \right).
\label{6d_cl}
\end{align}
Note that the moduli $\phi^I_{a}$ possess the mass dimension 2.
Then the second assumption that the leading contribution of the moduli  at a long distance is proportional to $(\partial \phi)^4$, which is the same as 
the D$p$-brane (\ref{action_one}) and M2-brane (\ref{ABJM_one}) cases, ensures 
\begin{align}
S_{\text{M5},T=0}^{\text{int}} \sim  -\int d \tau d^5x \sum_{a<b} \left(  \frac{ \{ \partial  (\phi_a -\phi_b) \partial (\phi_a -\phi_b)\}^2 }{|\phi_a -\phi_b|^{3}} \right) + \ldots\,,
\label{6d_one}
\end{align}
where the $|\phi_a -\phi_b|^{3}$ factor in the denominator is fixed by the dimensional analysis. This interaction is consistent with the gravitational potential between M5-branes in the eleven-dimensional spacetime.

The following discussion goes similarly to the previous cases.
By imposing the assumptions (\ref{phi-same}) and (\ref{kin-beta}),
the actions can be estimated as
\begin{align}
S_{\text{M5}}^{\text{kin}}  \sim  \int d^5 x \frac{N}{\beta } \phi^2\,, \quad
S_{\text{M5},T=0}^{\text{int}} \sim \int  d^5 x  \frac{N^2 \phi}{\beta^3 }\,.
\end{align}
By assuming $S_{\text{M5}}^{\text{kin}}  \sim S_{\text{M5},T=0}^{\text{int}}$, we determine the scale $\phi$ as
\begin{align}
\phi \sim \frac{N}{ \beta^2}\,.
\label{moduli-6d}
\end{align}
This is consistent with the dual gravity (\ref{size-M5})
and also with the suppression condition $\beta\sqrt{\phi}\gg 1$ at large $N$.
The free energy is finally estimated as
\begin{align}
F_{\text{M5}} \sim S_{\text{M5}}/\beta \sim N^3  T^6 V_5\,.
\end{align}
This result agrees with the dual gravity prediction (\ref{F-M5}).

Let us shortly comment on the equivalence of 5d SYM and 6d $\cN=(2,0)$ theory, which has been recently proposed~\cite{Douglas:2012,Lambert:2012}.\footnote{
Some problems on the 5d/6d equivalence have been reported in \cite{Bern:2012di}.}
For D4-branes, by using the Smilga-Wiseman method, 
we obtain the results (\ref{phi-scale}) and (\ref{F-Dp}), that is,
\ba
\phi_\text{D4}\sim \lambda T^2\,,\quad
F_\text{D4}\sim \lambda N^2 T^6 V_4 \,.
\ea
Since $\lambda \sim R_{11} N$, where $R_{11}$ is the radius of the M-theory circle, the following relations are satisfied:
\ba
\phi_\text{D4} \sim R_{11} NT^2 \sim R_{11} \phi_\text{M5}\,,\quad
F_\text{D4} \sim  N^3 T^6 R_{11} V_4 \sim F_\text{M5}\,.
\label{M5-D4}
\ea
These are consistent with the 5d/6d equivalence.\footnote{
Note that M5-branes in this case are wrapping on the M-theory circle, so Gregory-Laflamme transition like in the M2/D2 case may not occur.}
($\phi_\text{D4} \sim R_{11} \phi_\text{M5}$ is the natural identification as used in {\em e.g.}~\cite{Ho:2008}.)
We should emphasize that all these relations are explained through the field theory analyses.

To summarize, we obtain the free energy of $N$ M5-branes which reproduces the supergravity result.
This means that our assumptions for the effective M5-brane theory are not widely wrong.

\section{What was the $O(N^2)$ puzzle? }
\label{sec:argue}

In the D$p$-brane theories, 
the $O(N^2)$ free energies have been predicted both from 
weak and strong coupling regime.
On the other hand, in the M-brane theories,
 the different power of the $N$-dependence of the free energy has been predicted from the gravity, and some people believed that the dynamics of the M-branes
is quite distinct from the D-branes.
In this section, we argue whether M-branes really show distinct dynamics.

Let us recall why the $O(N^2)$ free energy is expected in the D-brane theories at
weak coupling.
There, the kinetic term will dominate and  all the fields are light, and the free energy becomes $O(N^2)$.

On the other hand, the Smilga-Wiseman method assumes that the dynamics of the classical moduli dominates at 
strong coupling
and $\beta |\phi_a-\phi_b|^{\xi/2} \gg 1$ is satisfied.
(Here $\xi=2$ for the D-branes, 4 for the M2 and 1 for the M5-brane, as argued in appendix~\ref{sec:app}.)
Then only the moduli are classically light and we might naively expect that the free energy becomes $O(N)$.
However the non-trivial $N$ dependences are induced from the interaction (\ref{action_one}), (\ref{ABJM_one}) and (\ref{6d_one}).
Thus even the appearance of the $O(N^2)$ free energy of the D-branes is non-trivial in this method!

The origin of the $O(N^2)$ free energy of the D-branes is the 't Hooft limit (fixed  $\lambda=g_{YM}^2 N $) as we can read off from eq.\,(\ref{F-Dp}).
This reason is quite different from the appearance of the $O(N^2)$ free energy at weak coupling.\footnote{Although the dynamics at
strong coupling and weak coupling
is different in the D$p$-brane theories, they would be smoothly connected \cite{Itzhaki:1998dd, Wiseman:2013, Anagnostopoulos:2007fw}.
If the coupling becomes so strong
that M-theory or S-dual description is required, transitions would happen in some cases \cite{Itzhaki:1998dd}.  }

On the other hand, we do not take the 't Hooft limit in the M-brane theories, and indeed we obtain the different $N$-dependence in the free energies.

These results suggest that it is not correct that 
the dynamics of M-branes are distinct.
Both D and M-branes would exhibit similarly behaviors at 
strong coupling
where the classical moduli would dominate.
A special thing for the D-branes is that the free energy at the 't Hooft limit becomes $O(N^2)$ similar to the free energy at 
weak coupling.
Therefore we conclude that the dynamics of M-branes is not special but the 't Hooft limit is special.
The smooth connections between the M2 and D2-branes (\ref{M2-D2}) and between the M5 and D4-branes (\ref{M5-D4}) also support this point.

\section{Discussions}
\label{sec:Disc}

In this article, we derived the thermodynamics of the $N$ M2 and
M5-branes 
by using the Smilga-Wiseman method.
All the results agree with the supergravity predictions through the
AdS/CFT correspondence.
Thus, although the Smilga-Wiseman method requires several assumptions
like eq.\,(\ref{kin-beta}) and the justification has not been done, we
believe that this method captures 
the dynamics of the branes and it would provide
some clues to prove the
gauge/gravity correspondence.

Especially this method suggests that the temperature independent
one-loop potential at a long distance, which behaves as $(\partial
\phi)^4/|\phi|^{7-p}$ in the D$p$-brane cases and $(\partial
\phi)^4/|\phi|^{8-p}$ in the M-brane cases, is important to reproduce
the gravity results.
Since this potential is almost uniquely determined by the
supersymmetry, the agreement between the gauge theory and gravity
might be a consequence of some kinematics.

In this method, we consider only the classical and one-loop parts of the effective action. However, 
Smilga has pointed out that, at the scale of the order (\ref{phi-scale}),
 the higher loop corrections appear to be the same order as the
classical and one-loop potential in the D0-brane case \cite{Smilga:2008}.
Thus we should understand why the gravity results are reproduced without considering the higher loop corrections.
A hint may be in the bosonic BFSS model.
Smilga has estimated the free energy of this model just by evaluating the classical and one-loop potential,\footnote{Different from the supersymmetric BFSS case (D0-brane case), the classical moduli is not relevant in the bosonic BFSS model.} although the higher loop corrections are the same order similar to the D0-brane case \cite{Smilga:2008}.
Interestingly this result agrees with an analytic calculation through the $1/D$-expansion, in which all higher-loop corrections are considered \cite{Mandal:2009vz}.
Hence we naively expect that the calculation up to the one-loop order might be enough to estimate the order of the free energy in strong coupling gauge theories for some mechanism.
Wiseman has argued that it might be due to the generalized conformal
symmetry \cite{Jevicki:1998ub} in the case of the D$p$-brane system
\cite{Wiseman:2013}.
If so, our discussion in the M-brane system suggests that the superconformal symmetry must play the same role.

From the perspective of M-theory, some people might be
disappointed at our results, since the mysterious $N$-dependence of the free energies of 
$N$ M2 and M5-branes is explained just in a similar way to the D$p$-branes
rather than a dramatic manner by using some characteristic natures of
the M-branes.
However, we should recognize that our work may provide 
another approach to clarify various aspects of M-brane dynamics.
Especially we find a new way to determine the multiple
M5-brane action, as argued in appendix~\ref{sec:app}.
For another aspect,
we should at least figure out the role of the superconformal symmetry in the dynamics.
We have obtained several hints on this problem from this work,
so it would be an interesting future work.

\sss{Acknowledgment}
We would like to thank to Masanori Hanada, Satoshi Iso, Yoshihisa Kitazawa, Yoshinori Matsuo, Jun Nishimura, Shigeki Sugimoto and Toby Wiseman for useful discussions and comments.
The work of T.M. is supported in part by Grant-in-Aid for Scientific Research (No. 24840046) from JSPS.
The work of S.S. is partially supported by Grant-in-Aid for JSPS fellows (No. 23-7749).

\appendix

\section{Predictions from supergravity}
\label{sec:gr}

Here we list the predictions for the black branes from supergravity.
We omit the precise coefficients which are not necessary in our arguments.
(See the references.) 

\paragraph{D$p$-brane system}

The free energy of a black D$p$-brane is \cite{Wiseman:2013}
\begin{align}
F \sim N^2 T^{\frac{2(7-p)}{5-p}} \lambda^{-\frac{3-p}{5-p}} V_p\,.
\label{F-Dp-gr}
\end{align}
This result is reliable when $N^{-\frac{2}{7-p}} \ll  (T \lambda^{-\frac{1}{3-p}} )^{\frac{3-p}{5-p}} \ll 1$ is satisfied.
The typical scale for the scalars would be given by the horizon size of the black brane \cite{Wiseman:2013}, 
\begin{align}
|\phi|  \sim T^{\frac{2}{5-p}} \lambda^{\frac{1}{5-p}}\,.
\label{size-Dp}
\end{align}

\paragraph{M2-brane system}

The metric of a black M2-brane in the Euclidean signature is
\begin{align}
ds^2 = &H(r)^{-2/3} \left( f(r) d\tau^2 + dx_1^2+ dx_2^2 \right) + H(r)^{1/3} \left( \frac{dr^2}{f(r)}+r^2 ds^2_{S^7} \right) ,\nt
&H(r)=1+ \frac{R^6}{r^6}\,,
 \qquad f(r)=1- \frac{r_0^6}{r^6}\,,
\end{align}
where $R \sim  N^{1/6} l_p$ is the radius of $AdS_4$ and $S^7$, and  $l_p$ is the eleven-dimensional Planck length.\footnote{We use a notation in \cite{Herzog:2002fn}.}
$r_0$ is the location of the horizon and
it is related to the Hawking temperature
through the smoothness condition at the horizon,
\begin{align}
T= \frac{3}{2\pi} \frac{r_0^2}{R^3}\,.
\end{align}
Here we have taken the near horizon limit $r \ll R$.
In order to consider the ABJM theory, we replace $S^7$ to $S^7/{\bbZ_k}$ and $R$ to $R' \sim (k N )^{1/6}l_p$ \cite{Aharony:2008}.

Now we evaluate the scale for the scalars $\phi$ and the free energy of  the ABJM theory.
Since the scale for the scalars is related to the location of the horizon $r_0$, we obtain
\begin{align}
|\phi| \sim \frac{r_0}{k^{\frac12}l_p^{\frac32}} \sim \frac{R^{\frac32}}{k^{\frac12}\beta^{\frac12}l_p^{\frac32}} \sim  \frac{N^{\frac14}}{ k^{\frac14}\beta^{\frac12}}\,.
\label{size-M2}
\end{align}
Here we have used the fact that the ABJM action (\ref{action-ABJM}) has the overall factor $k$, so we can read off the relation $(\text{length})^2/l_p^3 \sim k|\phi|^2$ from the kinetic term of the scalar fields.
Finally the free energy is obtained as \cite{Aharony:2008}
\begin{align}
F \sim N^{\frac32} \sqrt{k} T^3 V_2\,.
\label{F-ABJM}
\end{align}

\paragraph{M5-brane system}

The metric of a black M5-brane in the Euclidean signature is
\begin{align}
ds^2 = &H(r)^{-1/3} \left( f(r) d\tau^2 + \sum_{i=1}^{5} dx_i^2 \right) + H(r)^{2/3} \left( \frac{dr^2}{f(r)}+r^2 d \Omega_4^2 \right) ,\nt
&H(r)=1+ \frac{R^3}{r^3}\,,
 \qquad f(r)=1- \frac{r_0^3}{r^3}\,,
\end{align}
where $R \sim N^{1/3} l_p$ is the $AdS_7$ and $S^4$ radius.
$r_0$ is the location of the horizon is related to the Hawking temperature as,
\begin{align}
T= \frac{3}{4\pi} \frac{r_0^{1/2}}{R^{3/2}}\,.
\end{align}
Here we have taken the near horizon limit $r \ll R$.
Since the horizon $r_0$ is related to the scale for the scalars, we obtain
\begin{align}
|\phi| \sim \frac{r_0}{l_p^{3}} \sim \frac{R^{3}}{l_p^{3} \beta^{2}} \sim  \frac{N}{ \beta^{2}}\,.
\label{size-M5}
\end{align}
The free energy is given by
\begin{align}
F \sim N^{3}  T^6 V_5\,.
\label{F-M5}
\end{align}

\section{On one-loop corrections}
\label{sec:app}
In this appendix, we consider the one-loop corrections of the moduli effective actions.
In order to calculate the corrections in D$p$ and M2-branes~\cite{Wiseman:2013}, we consider the off-diagonal fluctuations around the classical solution (\ref{Dp-vacua}):
\ba
A_{\mu,ab}=a_{\mu,a}\delta_{ab}+\delta A^\mu_{ab}\,,\quad
\Phi^I_{ab}=\phi^I_a\delta_{ab}+\delta \Phi^I_{ab}\,,\quad
\Psi_{ab}=\delta \Psi_{ab}\,.
\label{logdet}
\ea
In addition to these fields, we need to introduce Fadeev-Popov ghost fields due to the gauge invariance.
There are no linear terms of fluctuations in the effective action since we are expanding around a classical solution, so the leading contribution comes from the quadratic fluctuations.
The contribution from the quadratic fluctuation of each field in the effective action can be written as
\ba\label{Dp-Delta}
\sum_n\sum_{a<b}\Tr'\left[\ln\left(-\partial^i\partial_i+\left(\frac{2\pi n}{\beta}\right)^2+|\phi_a-\phi_b|^{\xi}+\ldots\right)\right],
\ea
where $\Tr'$ means the trace over the spatial momenta, and $i$ denotes the spatial coordinates. For the summation, $n\in\bbZ$ for bosonic fields and $n\in\bbZ+1/2$ for fermionic fields.
In the D$p$-brane case, $\xi$ equals two. This power comes from the quadratic fluctuations of the four-scalar interactions in SYM theory.
In the M2-brane case, this power $\xi$ becomes four~\cite{Baek:2008},
which comes from the six-scalar interactions in the ABJM theory.
In our discussion only the scalar moduli fields $\phi$ play relevant roles, so here we omit the contribution from the gauge moduli $a_{\mu,a}$. 
After summing up the contributions from all the fields, we obtain the one-loop effective action as~\cite{Wiseman:2013, Baek:2008, Aharony:2005ew}
\ba
S^\text{one-loop} \sim -\int d\tau d^px \sum_{a<b}\left[\frac{(\partial\phi_{ab}\partial\phi_{ab})^2}{|\phi_{ab}|^{d-p-3}}+\ldots+e^{-\beta|\phi_{ab}|^{\xi/2}}(\ldots)\right],
\label{eff-one-loop-gen}
\ea
for $\beta|\phi_{ab}|^{\xi/2}\gg 1$,
where $\phi_{ab}:=\phi_a-\phi_b$ and $d$ is the dimension of whole spacetime, i.e. $d=10$ for D$p$ and 11 for M2-brane.
Here the temperature dependent terms\footnote{The temperature dependent terms of the M2-brane case is derived from eq.\,(\ref{logdet}) by applying calculations shown in appendix~A in \cite{Aharony:2005ew}.} are proportional to $\exp(-\beta|\phi_a-\phi_b|^{\xi/2})$  and suppressed when $\beta|\phi_a-\phi_b|^{\xi/2} \gg 1$.
This is an important point in the thermodynamics of the branes at strong coupling.

Now we discuss the M5-brane case.
Although the Lagrangian of the M5-branes is unknown, we presume that the fields which are not moduli would have masses proportional to $|\phi_a-\phi_b|$, since the single M5-brane action should be reproduced from the moduli fields, if we take $|\phi_a-\phi_b|$ sufficiently large.
It implies that the thermal corrections for the effective potential of the moduli fields, which are induced from the path-integral of these non-moduli modes, would be proportional to $e^{-\beta|\phi_{ab}|^{1/2}}$  through a dimensional analysis.
Therefore, without knowing the details of the 6d $\cN=(2,0)$ theory, we can estimate that the effective potential of the moduli $\phi_b^I$ for large $|\phi_a-\phi_b|$ is given by eq.~(\ref{eff-one-loop-gen}) with $d=11$ and $\xi=1$.
(The first term in eq.~(\ref{eff-one-loop-gen}) is speculated in eq.~(\ref{6d_one}).)
Note that different from the D$p$ and M2-brane cases, it is unclear whether this potential is derived from the one-loop computation.

\paragraph{Exploring the 6d $\cN=(2,0)$ theory  }

Suppose that the 6d $\cN=(2,0)$ theory is described by matrix type fields $\Phi^I_{ab}$ and the moduli $\phi^I_a$ are the diagonal components of them,\footnote{
The formulation of multiple M5-branes with Lie 3-algebra has been proposed in~\cite{Lambert:2010} and studied {\em e.g.} in~\cite{Honma:2011,Kawamoto:2011}, 
where the scalar fields have `fundamental' representations of this algebra.
However this model is incomplete in that the action cannot be written down,
so it should be not evidence to deny the scalar fields with matrix representations.
} what we can speculate from the above estimation of the effective potential (\ref{eff-one-loop-gen})?
Naively this potential may be derived from eq.~(\ref{Dp-Delta}) with $\xi=1$, which is the one-loop determinant of the massive fields including the off-diagonal components of $\Phi$.
If so, the theory may have the three-scalar interactions $\sim \Phi^3$.
However, the $\Phi^3$ term with transverse $SO(5)$ symmetry cannot exist.
It might suggest that the action of M5-brane theory is not the power series of the fields.
For example, the interaction term $\sim\sqrt{\Phi^6}$ can be written down, 
and its quadratic fluctuation can be $|\phi_a-\phi_b|(\delta \Phi)^2$, which reproduces the factor (\ref{logdet}) in the one-loop determinant.
In this way, our discussion may provide new clues to the description of the low energy effective theory of multiple M5-branes.


\begin{thebibliography}{99}

\bibitem{Klebanov:1996}
I. R. Klebanov and A. A. Tseytlin,
``Entropy of near-extremal black $p$-branes,''
Nucl. Phys. {\bf B475} (1996) 164
[arXiv:hep-th/9604089].

\bibitem{Itzhaki:1998dd}
  N.~Itzhaki, J.~M.~Maldacena, J.~Sonnenschein and S.~Yankielowicz,
  ``Supergravity and the large N limit of theories with sixteen
  supercharges,''
  Phys.\ Rev.\  D {\bf 58} (1998) 046004
  [arXiv:hep-th/9802042].






\bibitem{Wiseman:2013}
T. Wiseman,
``On black hole thermodynamics from super Yang-Mills,''
arXiv:1304.3938 [hep-th].

\bibitem{Smilga:2008}
A. V. Smilga,
``Comments on thermodynamics of supersymmetric matrix models,''
Nucl. Phys. {\bf B818} (2009) 101
[arXiv:0812.4753 [hep-th]].


\bibitem{Anagnostopoulos:2007fw} 
  K.~N.~Anagnostopoulos, M.~Hanada, J.~Nishimura and S.~Takeuchi,
 ``Monte Carlo studies of supersymmetric matrix quantum mechanics with sixteen supercharges at finite temperature,''
  Phys.\ Rev.\ Lett.\  {\bf 100}, 021601 (2008)
  [arXiv:0707.4454 [hep-th]].

\bibitem{Aharony:2004ig}
  O.~Aharony, J.~Marsano, S.~Minwalla and T.~Wiseman,
  ``Black hole-black string phase transitions in thermal 1+1-dimensional supersymmetric Yang-Mills theory on a circle,''
  Class.\ Quant.\ Grav.\  {\bf 21}, 5169 (2004)
  [arXiv:hep-th/0406210].

\bibitem{Kawahara:2007fn} 
  N.~Kawahara, J.~Nishimura and S.~Takeuchi,
  ``Phase structure of matrix quantum mechanics at finite temperature,''
  JHEP {\bf 0710}, 097 (2007)
  [arXiv:0706.3517 [hep-th]].
 
\bibitem{Mandal:2009vz}
  G.~Mandal, M.~Mahato, T.~Morita,
  ``Phases of one dimensional large N gauge theory in a 1/D expansion,''
  JHEP {\bf 1002}, 034 (2010).
  [arXiv:0910.4526 [hep-th]].



 
 \bibitem{Aharony:2008}
O. Aharony, O. Bergman, D. L. Jafferis and J. Maldacena,
``$\cN=6$ superconformal Chern-Simons-matter theories, 
M2-branes and their gravity duals,''
JHEP {\bf 0810} (2008) 091 
[arXiv:0806.1218 [hep-th]].


\bibitem{Maldacena:1997re} 
  J.~M.~Maldacena,
  ``The Large N limit of superconformal field theories and supergravity,''
  Adv.\ Theor.\ Math.\ Phys.\  {\bf 2}, 231 (1998)
  [hep-th/9711200].

\bibitem{Baek:2008}
J.-H. Baek, S. Hyun, W. Jang and S.-H. Yi,
``Membrane Dynamics in Three dimensional $\cN = 6$ Supersymmetric Chern-Simons Theory,''
arXiv:0812.1772 [hep-th].



\bibitem{Gregory:1994}
R. Gregory and R. Laflamme,
``The instability of charged black strings and $p$-branes,''
Nucl. Phys. {\bf B428} (1994) 399
[arXiv:hep-th/9404071].

\bibitem{Martinec:1998ja} 
  E.~J.~Martinec and V.~Sahakian,
  ``Black holes and the superYang-Mills phase diagram. 2.,''
  Phys.\ Rev.\ D {\bf 59}, 124005 (1999)
  [hep-th/9810224].

\bibitem{Drukker:2010}
N. Drukker, M. Marino and P. Putrov,
``From weak to strong coupling in ABJM theory,''
Commun. Math. Phys. {\bf 306} (2011) 511
[arXiv:1007.3837 [hep-th]].

\bibitem{Fuji:2011}
H. Fuji, S. Hirano and S. Moriyama,
``Summing Up All Genus Free Energy of ABJM Matrix Model,''
JHEP {\bf 1108} (2011) 001
[arXiv:1106.4631 [hep-th]].

\bibitem{Hanada:2012}
M. Hanada, M. Honda, Y. Honma, J. Nishimura, S. Shiba and Y. Yoshida,
``Numerical studies of the ABJM theory for arbitrary N at arbitrary coupling constant,''
JHEP {\bf 1205} (2012) 121 
[arXiv:1202.5300 [hep-th]].







\bibitem{PST}
P. Pasti, D. P. Sorokin and M. Tonin,
``Covariant action for a $D = 11$ five-brane with the chiral field,''
Phys. Lett. {\bf B398} (1997) 41
[arXiv:hep-th/9701037].

\bibitem{Bandos:1997}
I. A. Bandos, K. Lechner, A. Nurmagambetov, P. Pasti, D. P. Sorokin and M. Tonin,
``Covariant action for the superfive-brane of M theory,''
Phys. Rev. Lett. {\bf 78} (1997) 4332
[arXiv:hep-th/9701149].

\bibitem{Cederwall:1998}
M. Cederwall, B. E. W. Nilsson and P. Sundell,
``An Action for the superfive-brane in $D = 11$ supergravity,''
JHEP {\bf 9804} (1998) 007
[arXiv:hep-th/9712059].

\bibitem{Douglas:2012}
M. R. Douglas,
``On $D=5$ super Yang-Mills theory and $(2,0)$ theory,''
JHEP {\bf 1102} (2011) 011
[arXiv:1012.2880 [hep-th]].

\bibitem{Lambert:2012}
N. Lambert, C. Papageorgakis and M. Schmidt-Sommerfeld,
``M5-Branes, D4-Branes and Quantum 5D super-Yang-Mills,''
JHEP {\bf 1101} (2011) 083 
[arXiv:1012.2882 [hep-th]].

\bibitem{Bern:2012di} 
Z.~Bern, J.~J.~Carrasco, L.~J.~Dixon, M.~R.~Douglas, M.~von Hippel and H.~Johansson,
``$D = 5$ maximally supersymmetric Yang-Mills theory diverges at six loops,''
  Phys. Rev. D {\bf 87} (2013) 025018
  [arXiv:1210.7709 [hep-th]].


\bibitem{Ho:2008}
P.-M. Ho, Y. Imamura, Y. Matsuo and S. Shiba,
``M5-brane in three-form flux and multiple M2-branes,''
JHEP {\bf 0808} (2008) 014 
[arXiv:0805.2898 [hep-th]].




\bibitem{Jevicki:1998ub} 
  A.~Jevicki, Y.~Kazama and T.~Yoneya,
  ``Generalized conformal symmetry in D-brane matrix models,''
  Phys.\ Rev.\ D {\bf 59}, 066001 (1999)
  [hep-th/9810146].


\bibitem{Herzog:2002fn} 
  C.~P.~Herzog,
  ``The Hydrodynamics of M theory,''
  JHEP {\bf 0212}, 026 (2002)
  [hep-th/0210126].


\bibitem{Aharony:2005ew} 
  O.~Aharony, J.~Marsano, S.~Minwalla, K.~Papadodimas, M.~Van Raamsdonk and T.~Wiseman,
  ``The Phase structure of low dimensional large N gauge theories on Tori,''
  JHEP {\bf 0601}, 140 (2006)
  [hep-th/0508077].



\bibitem{Lambert:2010}
N. Lambert and C. Papageorgakis,
``Nonabelian $(2,0)$ Tensor Multiplets and 3-algebras,''
JHEP {\bf 1008} (2010) 083
[arXiv:1007.2982 [hep-th]].

\bibitem{Honma:2011}
Y. Honma, M. Ogawa and S. Shiba,
``D$p$-branes, NS5-branes and U-duality from nonabelian $(2,0)$ theory with Lie 3-algebra,''
JHEP {\bf 1104} (2011) 117 
[arXiv:1103.1327 [hep-th]].

\bibitem{Kawamoto:2011}
S. Kawamoto, T. Takimi and D. Tomino,
``Branes from a non-Abelian $(2,0)$ tensor multiplet with 3-algebra,'' 
J. Phys. {\bf A44} (2011) 325402 
[arXiv:1103.1223 [hep-th]].




\end{thebibliography}
\end{document}